# SOXS mechanical integration and verification in Italy

M. Aliverti*a, F. Battaini[f], K. Radhakrishnan[f], M. Genoni[a], G. Pariani[a], L. Oggioni[a], O. Hershko[i], M. Colapietro[c], S. D'Orsi[c], A. Brucalassi[n], G. Pignata[n,o], H. Kuncarayakti[j,k], S. Campana[a], R. Claudi[f], P. Schipani[c], J. Achrén[l], J. A. Araiza Duran[m,o], I. Arcavi[p], A. Baruffolo[f], S. Ben Ami[i], R. Bruch[i], G. Capasso[c], E. Cappellaro[f], R. Cosentino[h], F. D'Alessio[b], P. D'Avanzo[a], M. Della Valle[c], M. De Pascale[f], R. Di Benedetto[d], A. Gal Yam[i], M. Hernandez[h], J. Kotilainen[j,k], M. Landoni[a], G. Li Causi[s], S. Mattila[k], M. Munari[d], M. Rappaport[i], K. E.M.A. Redaelli[a], D. Ricci[f], M. Riva[a], A. Rubin[i], B. Salasnich[f], S. Smartt[t], R. Zanmar Sanchez[d], S. Scuderi[e], M. Stritzinger[u], E. Ventura[h], F. Vitali[b], D. Young[t]. [a]INAF Osservatorio Astronomico di Brera, Via Bianchi 46, I-23807, Merate, Italy; [b]INAF Osservatorio Astronomico di Roma, Via Frascati 33, I-00078 M. Porzio Catone, Italy; [c]INAF Osservatorio Astronomico di Capodimonte, Sal. Moiariello 16, I-80131, Naples, Italy; [d]INAF Osservatorio Astrofisico di Catania, Via S. Sofia 78, I-95123 Catania, Italy; [e]INAF IASF Milano Via A. Corti, 12, I-20133 Milano, Italy; [f]INAF Osservatorio Astronomico di Padova, Vicolo dell'Osservatorio 5, I-35122, Padua, Italy; [g]ESO, Karl Schwarzschild Strasse 2, D-85748, Garching bei München, Germany; [h]FGG INAF, TNG, Rambla J.A. Fernandez Perez 7, E-38712 Brenã Baja (TF), Spain; [i]Weizmann Institute of Science, Herzl St 234, Rehovot, 7610001, Israel; [j]Finnish Centre for Astronomy with ESO (FINCA), FI-20014 University of Turku, Finland; [k]Tuorla Observatory, Dept. of Physics and Astronomy, FI-20014 University of Turku, Finland; [l]Incident Angle Oy, Capsiankatu 4 A 29, FI-20320 Turku, Finland; [m]Centro de Investigaciones en Optica A. C., 37150 León, Mexico; [n]Universidad Andres Bello, Avda. Republica 252, Santiago, Chile; [o]Millennium Institute of Astrophysics (MAS); [p]Tel Aviv University, Department of Astrophysics, 69978 Tel Aviv, Israel; [q]Dark Cosmology Centre, Juliane Maries Vej 30, DK 2100 Copenhagen, Denmark; [r]Aboa Space Research Oy, Tierankatu 4B, FI-20520 Turku, Finland; [s]INAF Istituto di Astrofisica e Planetologia Spaziali, Via Fosso del Cavaliere 100, I-00133, Roma, Italy; [t]Astrophysics Research Centre, Queen's University Belfast, Belfast, BT7 1NN, UK; [u]Aarhus University, Ny Munkegade 120, D-8000 Aarhus.

**ABSTRACT**

SOXS (SOn of X-Shooter) is a medium resolution (~4500) wide-band (0.35 - 2.0 µm) spectrograph which passed the Final Design Review in 2018. The instrument is in the final integration phase and it is planned to be installed at the NTT in La Silla by next year. It is mainly composed of five different optomechanical subsystems (Common Path, NIR spectrograph, UV-VIS spectrograph, Camera, and Calibration) and other mechanical subsystems (Interface flange, Platform, cable corotator, and cooling system).

A brief overview of the optomechanical subsystems is presented here as more details can be found in the specific proceedings while a more comprehensive discussion is dedicated to the other mechanical subsystems and the tools needed for the integration of the instrument.

Moreover, the results obtained during the acceptance of the various mechanical elements are presented together with the experiments performed to validate the functionality of the subsystems. Finally, the mechanical integration procedure is shown here, along with all the modifications applied to correct the typical problems happening in this phase.

**Keywords:** NTT, SOXS, mechanical design, alignment, integration

---

*matteo.aliverti@inaf.it; phone +39 347 1504637; http://www.brera.inaf.it/

# 1. INTRODUCTION

The SOXS instrument is a medium resolution spectrograph mainly developed for transients follow-up and in the procurement and integration phase. An overall view of the instrument can be found in [1], while this paper mainly focuses on the mechanical integration of the various subsystems and tools.

In section 2 the general layout is briefly shown. The subsystems are shown in section 3 and 4 where the 'optical' subsystems mounted on the flange are described in the first while the other subsystems and the tools are described in the latter. In section 5 the integration status and procedure are shown.

More information on the mechanical design can be found in [2] and [3].

# 2. GENERAL LAYOUT

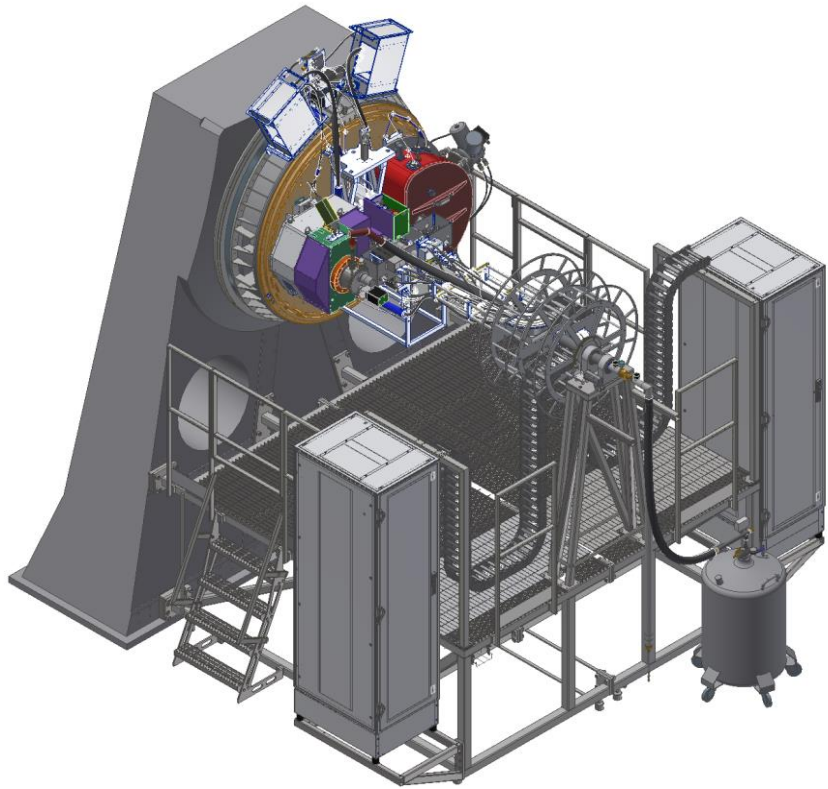

Figure 1. Overall view of the SOXS spectrograph.

The 3D model of the whole instrument can be seen in Figure 1. The rotating part of the NTT fork directly holds the interface flange (orange), the two NGCs for the UV-VIS and NIR detectors (2 light grey boxes on the top) and the cooling system holder (in between the two NGCs). The platform is connected to the non-rotating part of the NTT fork and sustain two electronic cabinets and the cable corotator. The LN2 line starts from the LN2 tank (only element on the Nasmyth floor), pass through the central hole of the cable corotator where a LN2 derotator is located.

# 3. MAIN INSTRUMENT

The core of the instrument is shown in Figure 2 and is composed by the "interface flange" (IF flange) holding the Common Path, the UV-VIS spectrograph and the NIR spectrograph via 3 sets of large kinematic mounts. Due to their low mass, the tight installation space, and the optical design, the Calibration box and the Acquisition Camera are connected to the Common Path structure via 2 sets of small kinematic mounts. As the mass and momentum limitations imposed by the telescope are quite tight, and to minimize thermal misalignments in operations, almost the entire instrument is manufactured in aluminium 6061/6082.

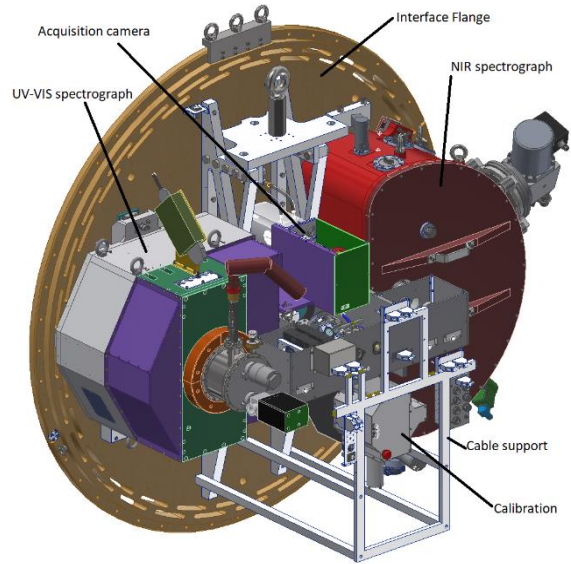

Figure 2. View of the interface flange and the 5 subsystems attached to it.

### 3.1 Interface flange

The interface flange's main role is to connect the spectrographs and the Common Path to the fork of the telescope. The element is composed by an aluminium 6082-T651 60mm thick flange with a central support and is shown in Figure 3.

On the pictures, different features can be noticed, including an M30 eyebolt used as central lifting element, three M12 eyebolts to move the interface flange alone, three sets of kinematic elements for the installation of the UV-VIS spectrograph (left side of the flange), the NIR spectrograph (right), and the common path (CP, centre).

Said Kinematic Mounts (KMs) have been aligned in Merate following the results of the alignment of the other side of the KM on the CP, UVVIS, and NIR subsystems. The results are shown in [3] but their positions will be adjusted during the integration of the various subsystem to obtain the desired coalignment between them.

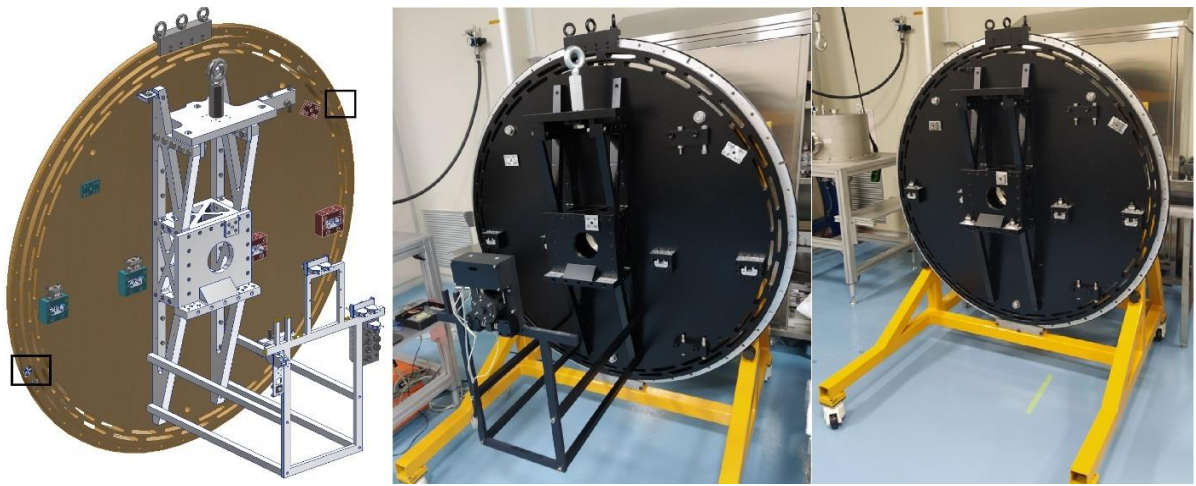

Figure 3. Left: 3D model of the IF flange with the regulation elements in the black boxes. Centre: photo of the IF flange with the cables support installed. Right: photo of the IF flange with the main kinemati mounts aligned.

Other useful elements are two special holes/slots used to increase repeatability of the flange and a series of alignment bars used to facilitate the installation of the two spectrographs (the four M20 holes for the NIR spectrograph are shown while the two rails for the UV-VIS spectrograph are not, as their installation is planned for a later stage).

In the central figure it is also shown the cables support which is used to install the disconnection points of the electrical cable to facilitate the dismounting of the instrument. Moreover, the corotator feedback is placed in the centre and its design is discussed in the corotator section.

## 3.2 Common Path (CP)

The overall view of the CP is shown in Figure 4. Following the light coming from the telescope, we have a shutter, the calibration stage, the camera stage and the dichroic. The reflected light goes to the UV-VIS spectrograph through a mirror, the ADC, a tip-tilt mirror and the UV-VIS pupil stop. The refracted light, instead, goes to the NIR spectrograph through a mirror, a tip-tilt mirror and a refocuser. Both the outputs have two baffle to avoid spurious light from entering the spectrographs. All those elements have been glued in Merate and, after a first check of the results, have been shipped to Padua together with the CP structure for the optical alignment of the subsystem (described in [4])

Other important elements are the kinematic mounts (in blue) used to install the CP on the interface flange and the Acquisition Camera on the CP. On the bottom part of the CP structure, another set of kinematic mounts identical to the acquisition camera one is used to install the calibration unit. The Thorlabs pinholes supports used for the pinhole alignment are indicated in green (a fifth one is behind the shutter) as for the Thorlabs magnetic supports used to mount a camera on the exit focal planes. The alignment of those elements is described in [3].

In the same figure, a picture of the completely installed CP without the optics to test the installation of all the components is shown together with the pre-aligned one. The first picture has been taken after the painting with Z306. The latter has all the optical elements installed and the pinholes and the dichroic aligned.

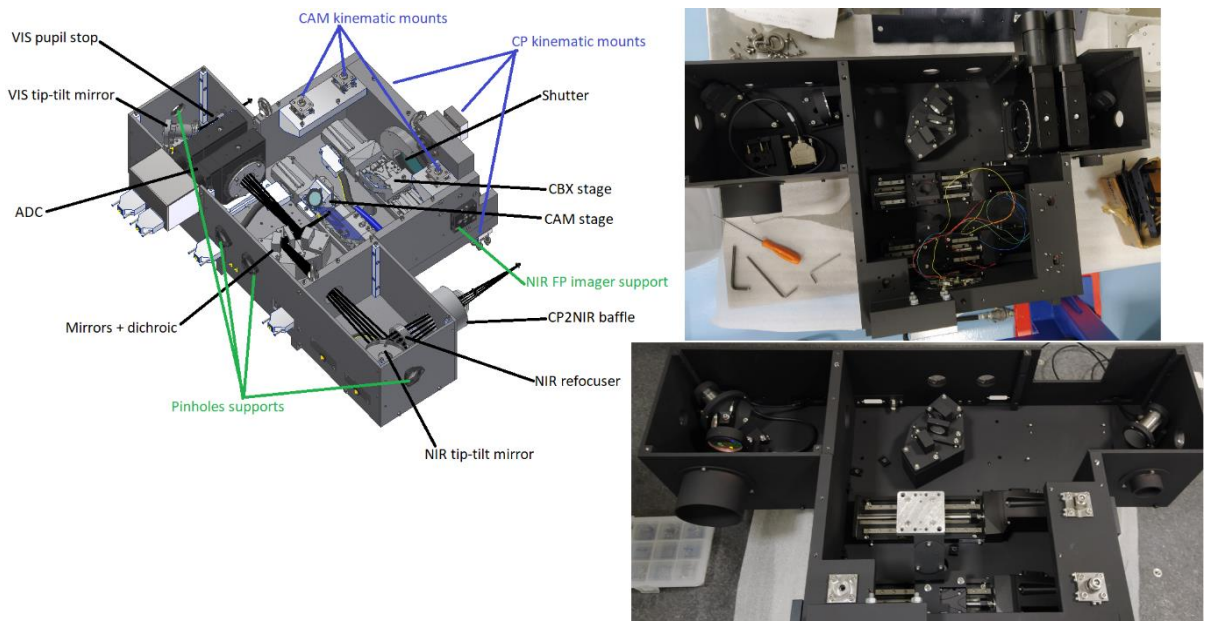

Figure 4. Overall view of the Common Path. Left: 3D model with the main components highlighted. Top right: picture with all the components installed. Bottom right: picture taken before the shipment to Padua for the final alignment.

The only missing optical elements in pictures are the ADCs and the holed mirror.

The ADC motors have been pre-aligned to their supports and shipped to the optical manufacturer together with the drawings of the optomechanical supports. The manufacturer had to glue the optical elements to the optomechanical supports, install them inside the motors, and ship the whole unit to Padua for the final integration. Due to some error in the gluing procedure the whole realignment has to be performed again in Padua and the whole process is described in [5].

The holed mirror is the camera selector (see **Error! Reference source not found.**) and has been received after the CP has been shipped to Padua.

The support has been manufactured and glued to the fused silica holed mirror. In particular they have been mounted together, pre-aligned via CMM using 5 push screws, and dismounted. After 6 RTV glue spot have been applied to the optomechanical support, the mirror has been placed again in position and realigned. After the curing a final check of the alignment has been done and 6 reference point have been taken to easy the final alignment inside the CP.

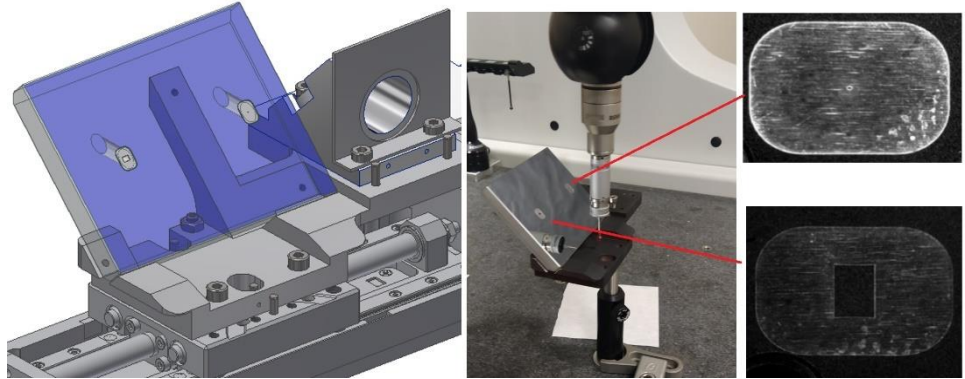

Figure 5. Camera selector mirror. Left: mounted on the camera selector stage. Centre: view of the element during the alignment. Right: pictures of the two holes (pinholes and scientific hole) mounted on the front face.

The deviations from the nominal position of the points are shown in Table 1. Note that the 'Focus' values define the focus and the tip-tilt, the 'Clock' values define the lateral displacement and the rotation around axis centred to the pinhole and perpendicular to the optical surface, and the 'Lateral' is just a placeholder as this position will be corrected by the stage.

Table 1. Deviation from the nominal positions of the holed mirror after the alignment.

| [mm] | Focus 1 | Focus 2 | Focus 3 | Clock 1 | Clock 2 | Lateral |
|---|---|---|---|---|---|---|
| Nominal | 0.2800 | 0.2800 | 0.2800 | 34.9800 | 34.9850 | -17.0180 |
| Actual | 0.2795 | 0.2797 | 0.2788 | 34.9709 | 34.9898 | -17.0180 |
| Deviation (A-N) | -0.0005 | -0.0003 | -0.0012 | -0.0091 | 0.0048 | 0.0000 |

Some detail on this mirror can also be found in [6].

### 3.3 Calibration Unit (CBX)

The calibration unit is composed by a 'light box' (centre in Figure 6) and an 'optic box' (left in Figure 6). The optic box is directly connected to the CP via three kinematic mounts. More details on this subsystem can be found in [8].

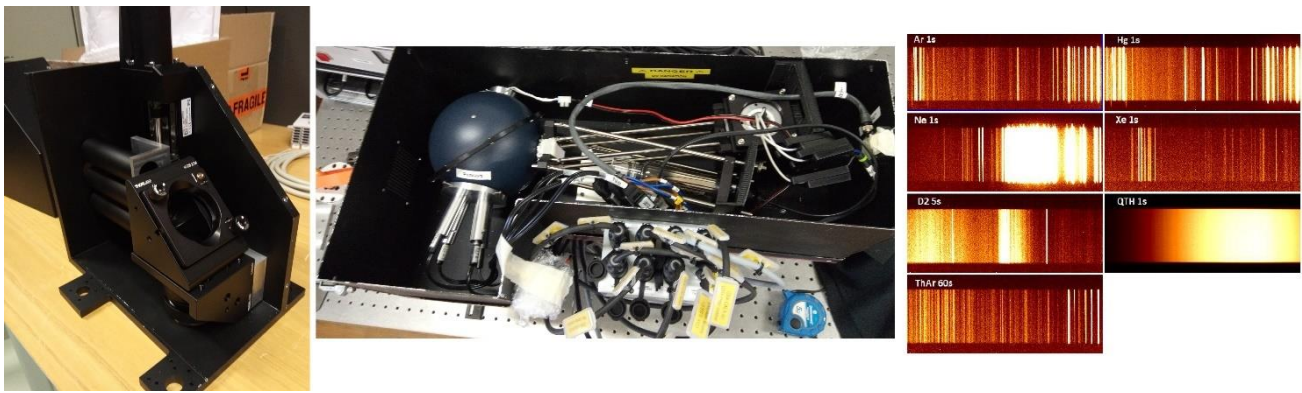

Figure 6. Left and centre: pictures of the calibration box. Right: spectra of the individual CBX lamps.

In Figure 6 it is also possible to see the spectra of the various lamps. If the penrays and the QTH require an integration time of about 1s to saturate the laboratory detector, the Deuterium requires 5 seconds and the ThAr reaches 20% of the saturation after 60s. This is related to the location of the last two lamps with respect to the integrating sphere.

A new design of the 'light box' part has been studied in order to move them close to the integrating sphere and facilitate the replacement of the lamps. This change required, as a consequence a large number of modifications, including a red button instead of the automatic switch and a new light box structure (see Figure 7).

The lamps have been moved in the following way:

- ThAr lamp directly mounted in a tube in front of one of the small apertures. The replacement is done dismounting the four screws on the back of the lamp.

- Deuterium lamp directly mounted in front of the other aperture. The replacement is done opening a small part of the cover and unscrewing 3 screws. Note that the collimating lens previously foreseen has been removed.

- The QTH is mounted in the large aperture together with the penrays. The replacement is dome unscrewing two screw and removing the small cover visible in the central picture of Figure 7.

- The penrays are mounted tangentially instead of radially and with a mask to reduce their efficiency and, therefore, to obtain a comparable flux between all the lamps. The expected flux reduction is between 50% and 85%.

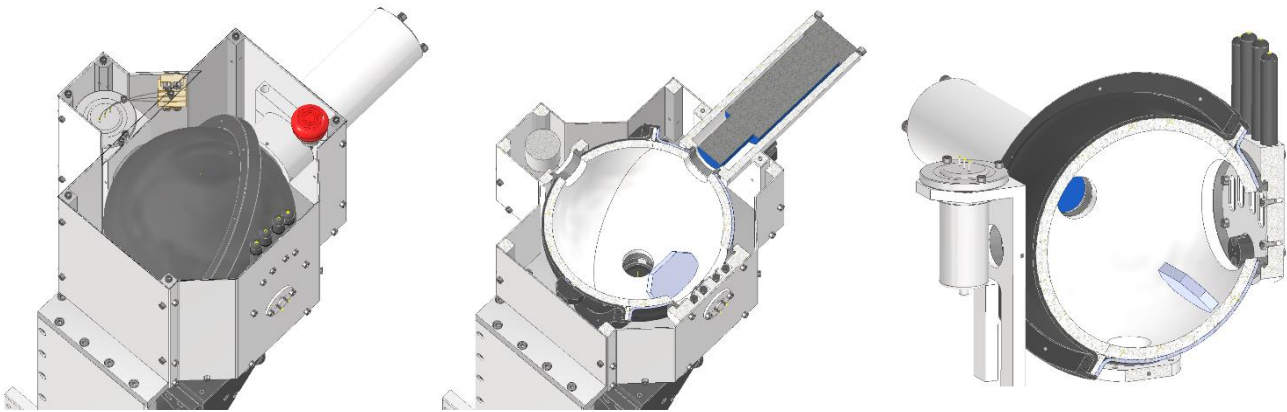

Figure 7. Modified design of the 'light box' part. Left: overall view. Centre: same view sectioned to the centre of the integrating sphere. Right: sections showing the penrays+QTH support.

The manufacturing of this element is ongoing and the test are expected in a few months, when the decision about which of the two to install will be taken.

### 3.4 Acquisition Camera (CAM)

The acquisition camera unit is shown in Figure 8. As in the CBX case, also this subsystem is connected to the CP via three kinematic mounts, which can be seen in the bottom of the figure. More details can be found in [6] and [9].

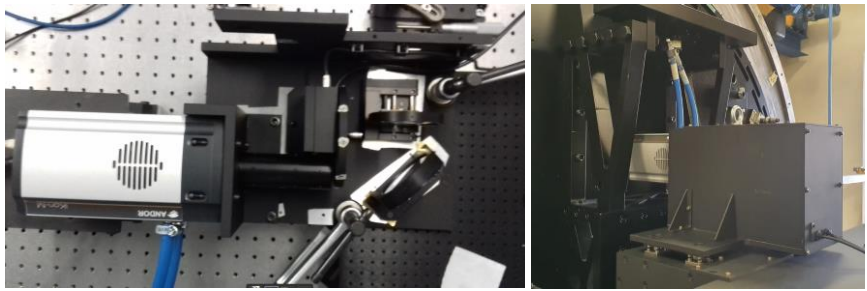

Figure 8. Left: acquisition camera during alignment. Right: acquisition camera mounted on the CP.

### 3.5 UV-VIS spectrograph

The UV-VIS spectrograph, shown in Figure 9, is directly connected to the interface flange via three large kinematic mounts. More details on this subsystem can be found in [10] while the detector part is shown in [11] and [12]. Details on the alignment procedure will be presented in a future proceeding.

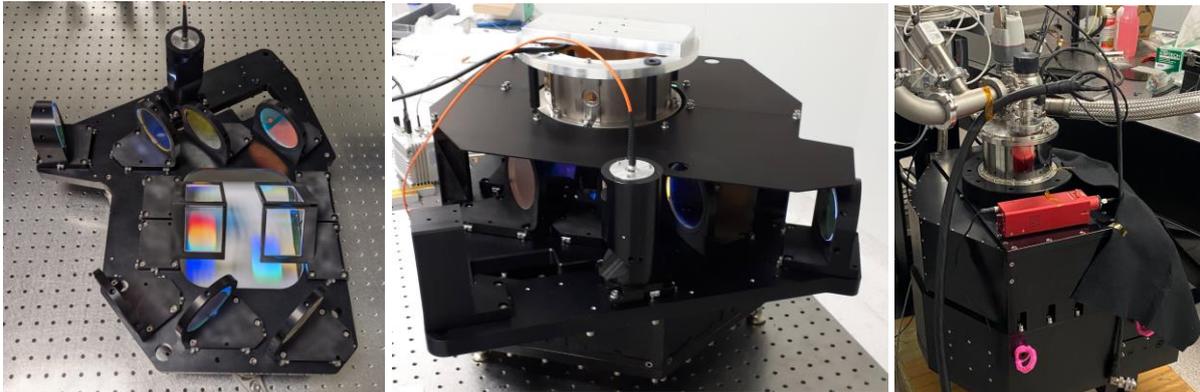

Figure 9. Left: top view of the feed part of the UV-VIS spectrograph. Centre: overall view of the opened spectrograph without the CFC. Right: overall view of the UV-VIS spectrograph.

### 3.6 NIR spectrograph

The NIR spectrograph is composed by a Vacuum Vessel with an internal bench mounted on inside it via four flexures shown in Figure 10 (left). On the bench 6 elements are mounted (following the light path):

- the 'slit sub-bench' which includes the slit, the pupil stop, the first lens and the first mirror;
- the main collimator spherical mirror;
- the 'dispersing sub-bench' which includes the collimator lens, the 3 prisms and the grating;
- the folding spherical mirror;
- the 'camera sub-bench' which includes the 3 camera lenses and the filter;
- the detector support.

Almost all the components in the subsystem are made of aluminium 6082 with the notable exceptions of the insulators (G10-FR4), the straps (copper and aluminium 1100), the flexures (titanium alloy) and the slit motor (titanium alloy).

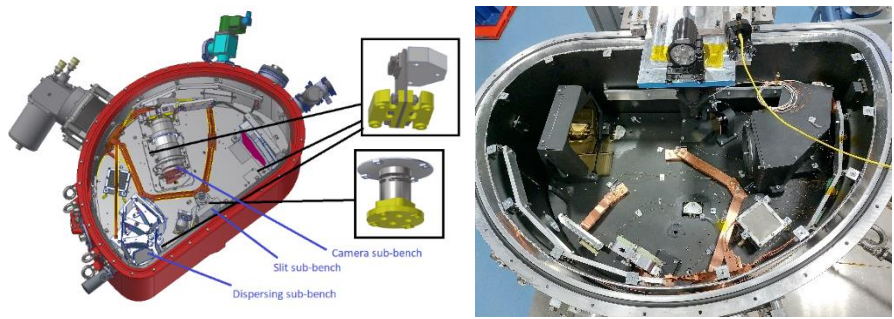

Figure 10. Left: overall view of the NIR spectrograph with the main sub-bench and the flexures location highlighted. Right: vacuum vessel after the 'hot alignment'.

The integration of the system has been preformed in a parallelized way. The cryogenic and vacuum tests, also described [13], have been done filling the vacuum vessel with dummy elements to simulate the thermal capacity of the real optomechanical elements (see Figure 11, left). The warm optical alignment has been performed on a dummy optical bench (see Figure 11, right) and will be described in a future proceeding.

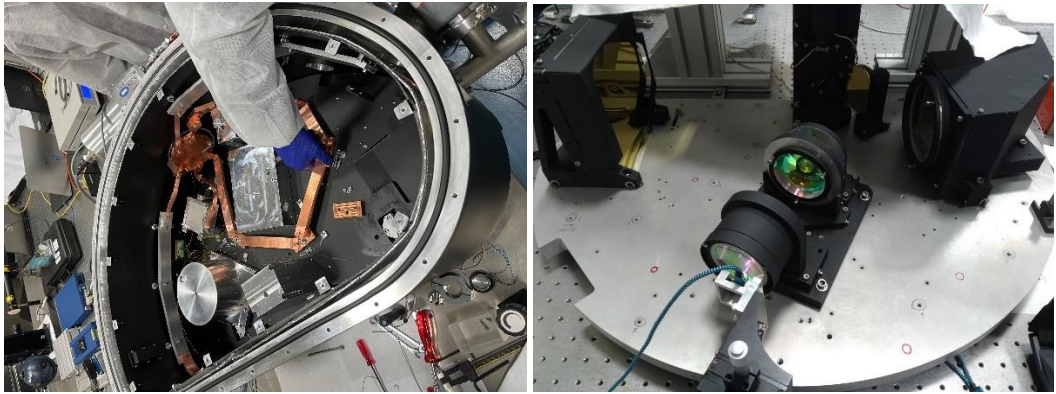

Figure 11. Left: Vacuum vessel with the dummy elements inside. Right: dummy bench with all the optomechanical elements (except for the NIR detector).

The slit sub-bench can be seen in Figure 12. This subsystem includes, from left to right, the baffle, the pupil stop, a small lens kept in position with 3 springs, the slit motor, another baffle and the first flat mirror.

All those elements have been aligned with respect to the references (bottom surface and location of the 3 pins) in order to minimize the alignment of the subsystem on the bench. The only optical alignment has been done on the slit supports where, using a telecentric camera, the slits have been placed parallel to the movement of the piezo stage.

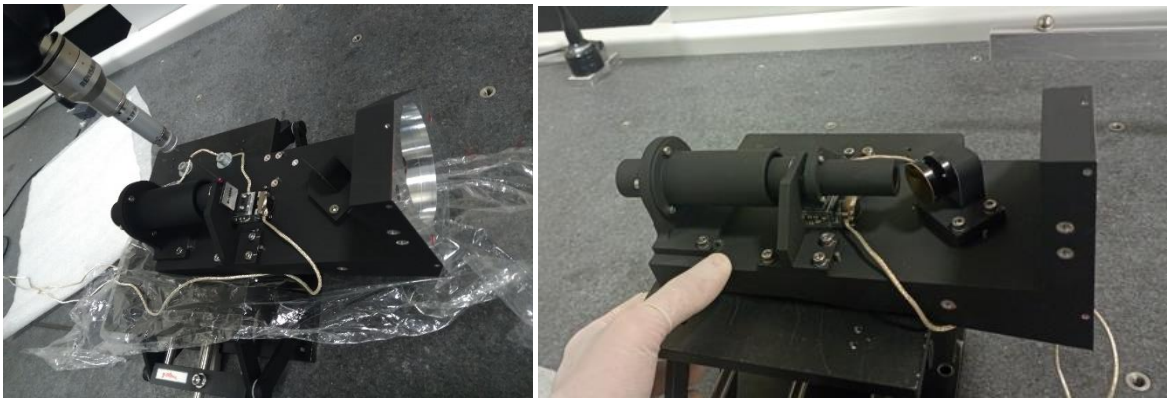

Figure 12. Slit sub-bench. Left: during the alignment of the slit support. Right: after the alignment.

The dispersing sub-bench (Figure 13) has been designed and built considering two possible configurations for the installation of the prisms:

1. with the 3 lateral fixed points placed one on one optical surface and the two on the other. Two springs are placed on the back of the prism and on one optical surface (Figure 13, left);
2. with the 3 lateral fixed points placed one on the back of the prism and the two on one optical surfaces. One spring is placed on remaining optical surface (Figure 13, right).

The first configuration maintains the same position of the prisms both in the warm and cold conditions but it could induce dangerous stresses in the elements in case the friction coefficient was not estimated correctly. Considering that the expected movement of the optical surfaces should not impact the optical quality, the second version has been chosen.

The position of all the prisms has been just verified via CMM as no regulation was possible. All the prisms have been placed with a focus error below 0.067mm and a tip-tilt error below 2 arcmin. The corrector lens has been placed in its nominal position as theoretical analysis showed that the direct aliment via CMM contacting the optical surfaces was less sensitive (see [14] for the method used).

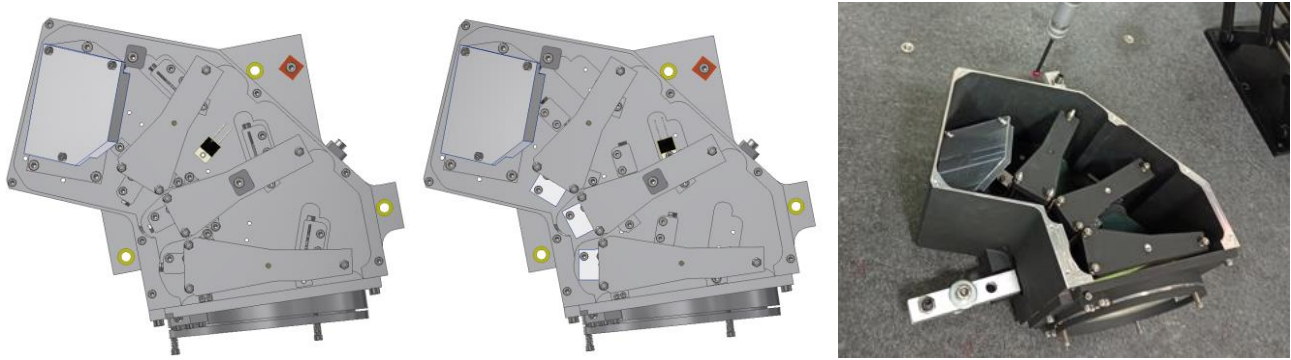

Figure 13. Dispersing sub-bench. Left: version 1. Centre: version 2. Right: dispersing sub-bench during the alignment.

The camera sub-bench (Figure 14) has been tested extensively. First, the optomechanical elements have been thermally cycled using a dedicated vacuum vessel and some dummy lenses to prove the repeatability and the safety of the conical springs system already presented in [2]. The bench without the optic has then been mechanically verified and aligned via CMM (Figure 14, left). Once the optical elements have been installed optical alignment with laser has been performed and the final positions have been checked via CMM (Figure 14, right). This final measure proved that a purely mechanical would have been sufficient.

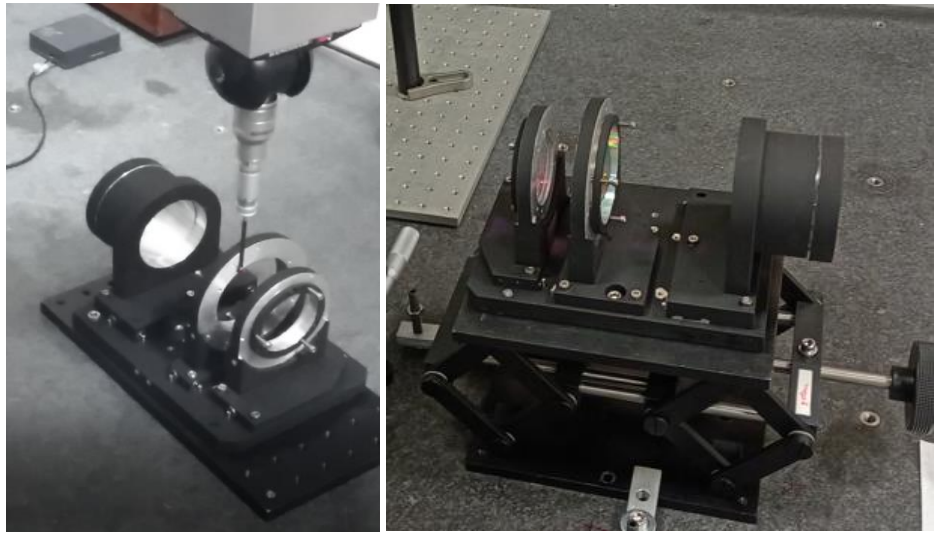

Figure 14. Left: camera sub-bench without the optical elements inside for the preliminary CMM alignment. Right: camera sub-bench with the optical elements inside for the final CMM check.

The aluminium mirrors are shown in (Figure 15). On the left the main collimator with three slots to reduce the stress on the mirror due to the screws. To furtherly reduce the stress and to compensate the rotation of the image on the detector due to gravity, 3 flexures have also been manufactured on the optomechanical support. In the central picture the first folding mirror with a wineglass foot and 3 slots, again, to reduce the stress due to the screws. And to loose the manufacturing tolerances of the mirror mounts. The same concept has been used for the last folding mirror.

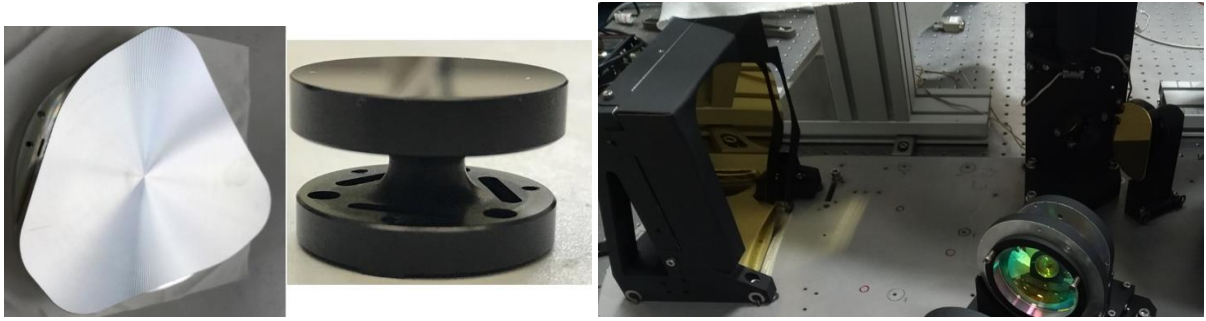

Figure 15. Left: aluminium mirrors under manufacturing. Right: aluminium mirrors installed on the dummy bench.

Finally yet importantly, the NIR telescope simulator, shown in Figure 16, has been manufactured and aligned. It is composed by a fibre support, two parabolic aluminium mirrors and a pupil stop. The telescope simulator has been mounted and aligned first on the dummy bench (Figure 16, right) and then on the NIR vacuum vessel cover for the final spectrograph alignment.

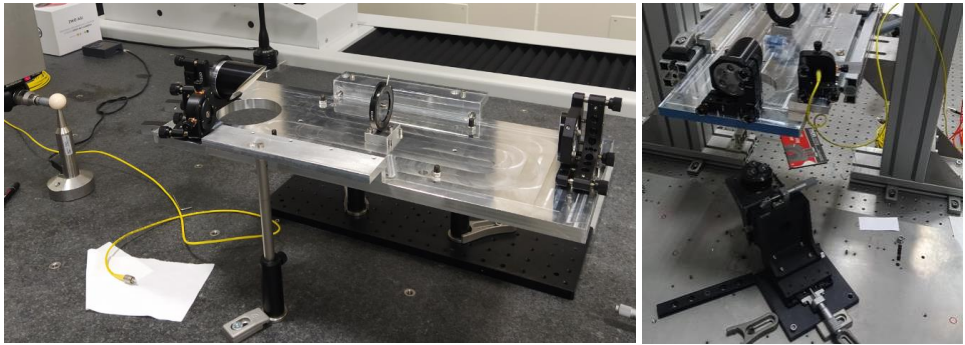

Figure 16. Telescope simulator system Left: during the internal alignment. Right: mounted on the NIR spectrograph for the warm alignment.

## 4. AUXILIARY ELEMENTS

In addition to the core of the instrument, other auxiliary elements have to be manufactured. Those include the platform, the corotator part, and the LN2 line.

### 4.1 Platform

The 3D design of the platform is shown in Figure 17. It is divided in a number of parts including two mirrored main bodies, a corotator support, two cable supports, two rack supports, 6 guardrails, 2 step ladders (each with 2 guardrails) and two doors. The part is currently under manufacturing (expected delivery January 2023) and three last modifications have been recently applied:

- the addition of cabinets dampers,
- the connection of the cabinets also from the holes above,
- the addition of two bars on the corotator support to avoid over-rotation of the LN2 line in case of failure of the LN2 derotator.

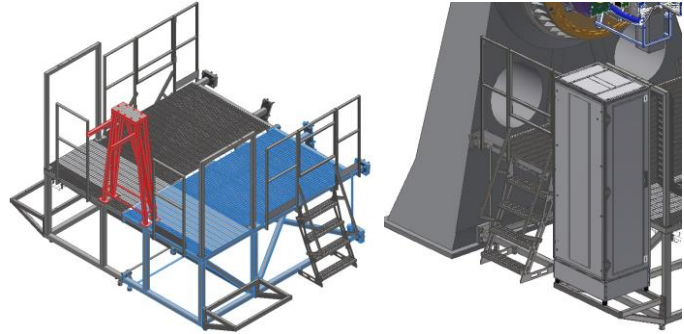

Figure 17. Left: Isometric view of the platform; in red the corotator support and in blue one of the two main parts. Right: detailed view showing the added cabinet dampers.

**4.2 Corotator and corotator feedback and LN2 line**

The corotator and the corotator feedback are shown in Figure 18. The corotator will be installed on the corotator support visible in Figure 17 (red part) using two spherical joints.

The corotator has been connected to the electronic control and, using two bellow couplings, to the corotator feedback. The system was able to maintain the correct rotational alignment between the two subsystems and, therefore, it has been shipped to Padua for the installation on the NTT simulator.

Before the shipment two additions related to safety have been applied: two protective plexiglass plates on the wheel and one 3D printed cover meant to avoid crushing hazards due to the gear.

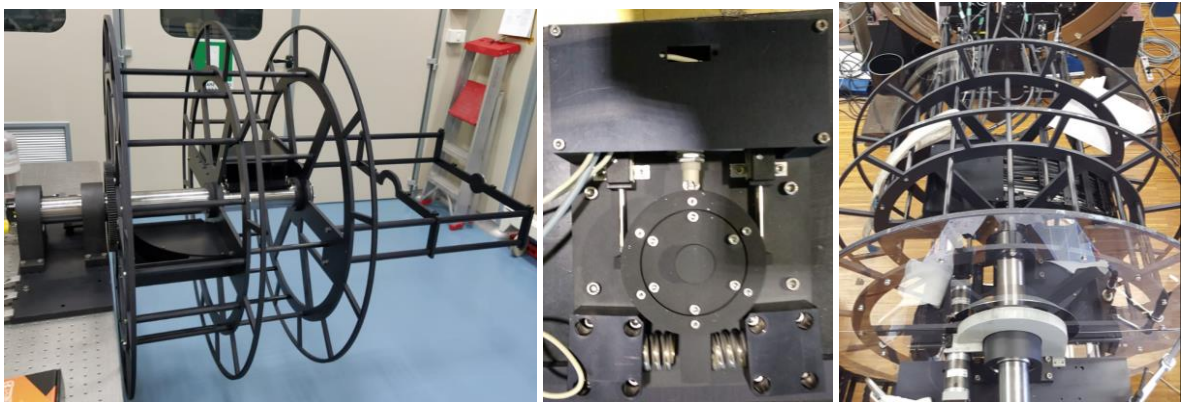

Figure 18. Left: overall view of the corotator. Centre: front view of the feedback system. Right: whole system with the protections mounted.

**4.3 LN2 line**

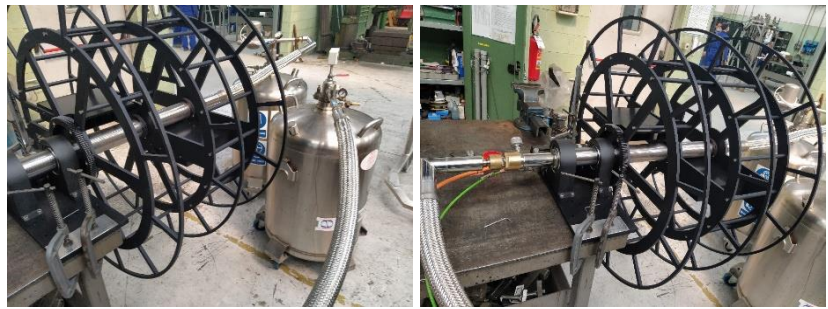

Figure 19. Overall view of the LN2 line during the acceptance tests at the manufacturer's factory.

The liquid nitrogen line, shown in Figure 19, is composed by 3 parts:

1. An input flexible line connecting the tank to the LN2 derotator
2. A rigid LN2 derotator
3. An output flexible line connecting the LN2 derotator to the CFC

The flexible line both follow a similar concept with male Johnston coupling at both ends (see Figure 20, left) and one of the ends tilted at 90° or 135° (see Figure 20, right).

There are minor differences between the various ends mainly in the shape and material of the element contacting the male and the female part and on the length of the male part itself. Those differences are due to the different constrains of the four different connections (in order from tank to CFC: ESO standard gravity invariant, relative rotation with valve, absence of valve, ESO standard gravity variant).

To avoid errors in the installation of the lines, each of them has an M35x1.5 long brass nut on one end and an M36x1.5 short brass nut on the other.

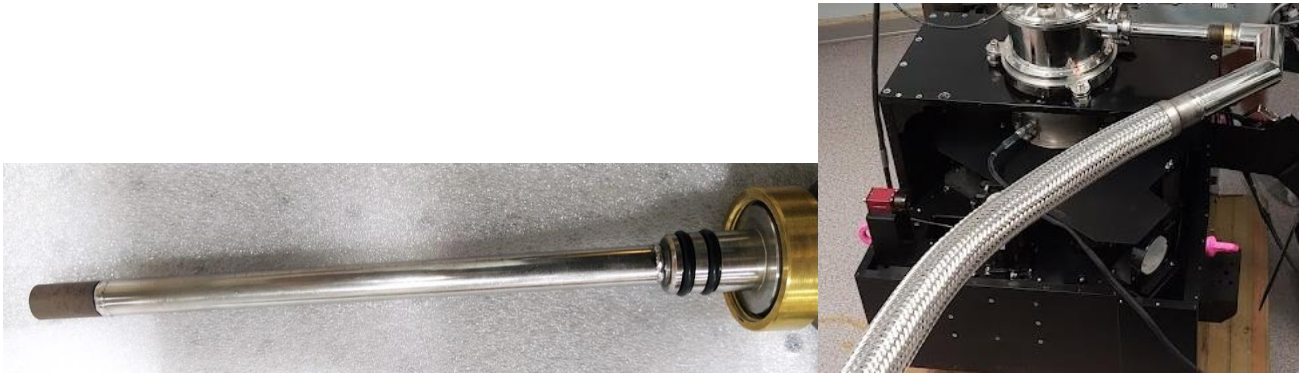

Figure 20. Example of the flexible lines. Left: male Johnston coupling. Right: output flexible line connected to the CFC.

The rigid LN2 derotator is a stainless steel vacuum tube with two female Johnston couplings. The one on the CFC side is a static one while the one on the tank side contains the rotating joint.

This element can be seen in Figure 21. A pressure gauge (first image on the left) is fitted between the first and the second O-ring (image on the right) to check if there is some leak on the first O-ring. A complete replacement of the O-rings and the oil seal is recommended at this point.

The maintenance is performed removing the brass nut, extracting the male rotating part and replacing all the wear out elements: the 2 O-rings 29.74x3.53 and the oil seal 30x40x7 on the female part and the bearings and the brass insulation on the male part.

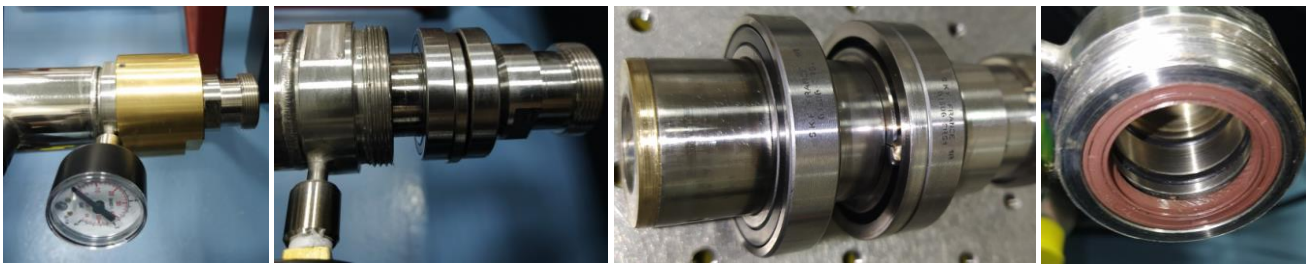

Figure 21. Disassembly of the rotating joint of the LN2 derotator. From left to right: entire element; after the brass nut removal; view of the male rotating part; view of the female part.

Another detail of the design is shown in Figure 22. The connection/disconnection is done tightening the M35x1.5 brass element but this operation can be performed only it the three radially mounted M5 screws are removed. When said screws are inserted the line can be extracted just enough to close the valve and avoid backflow of LN2 while exchanging the tank.

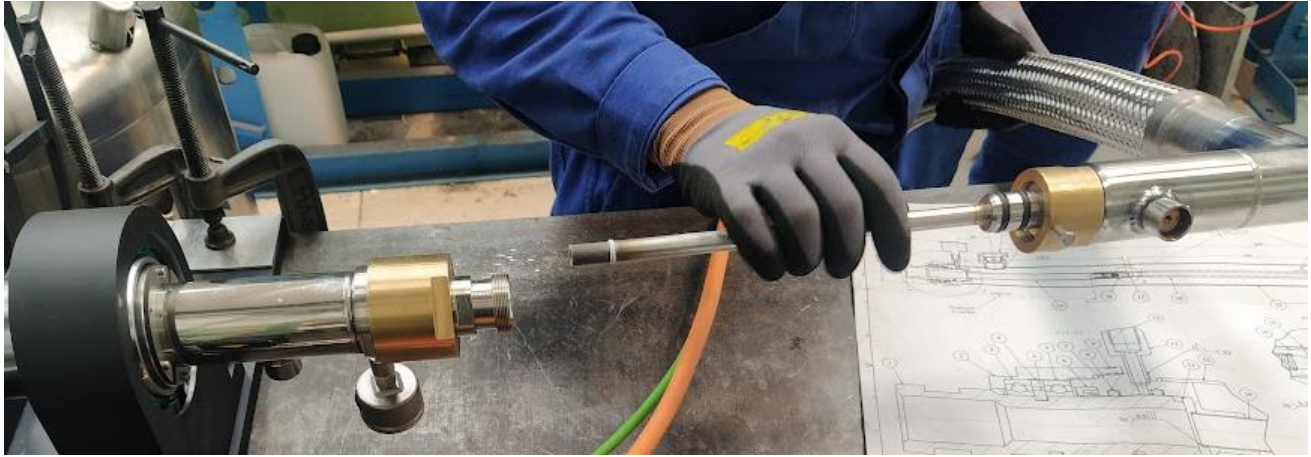

Figure 22. Installation of the flexible line on the rotating joint.

## 5. INTEGRATION

The preliminary integration of some subsystem has been performed Brera but most of them are occurring directly in Padua. A qualitative description of those operations is presented in the sext subsections

### 5.1 Preliminary installations in INAF Brera

The test installations performed in Brera consisted in mounting the IF flange on the SOXS support (see Figure 23) and the installation of the CP and the NIR spectrograph on the flange.

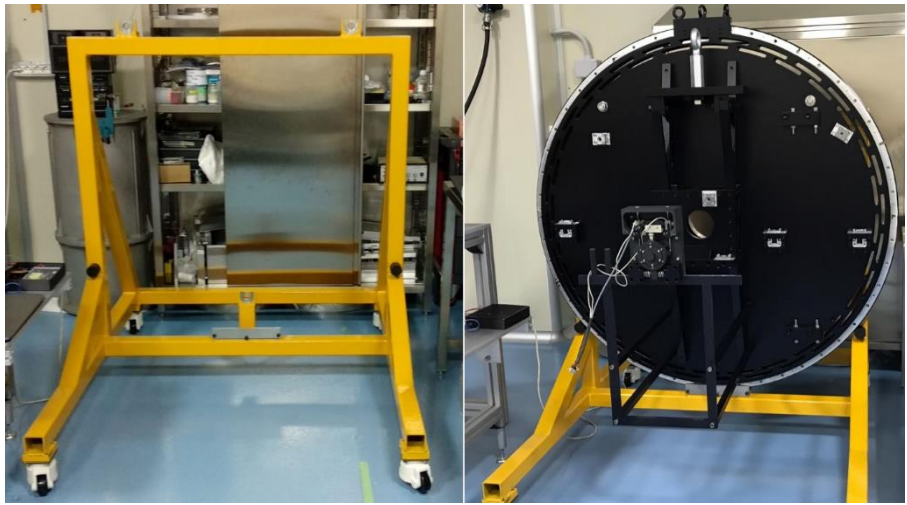

Figure 23. SOXS support (left) and IF installation (right).

The NIR spectrograph installation is shown in Figure 24. The spectrograph has been moved on the tilting table while being in the horizontal condition (used for maintenance and integration). The tilting table has then been rotated and the spectrograph lifted and mounted in the IF flange.

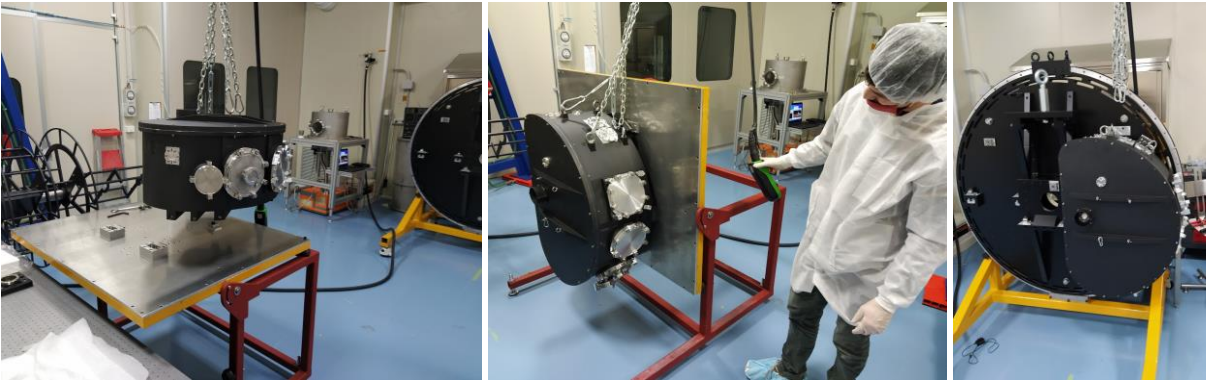

Figure 24. Left: installation of the NIR spectrograph on the tilting table. Centre: rotation on the tilting table. Right: NIR spectrograph installation on the IF flange.

The CP installation Figure 25. As the maintenance and installation conditions are identical, in this case there is no need to use the tilting table and the subsystem si directly mounted on the IF flange.

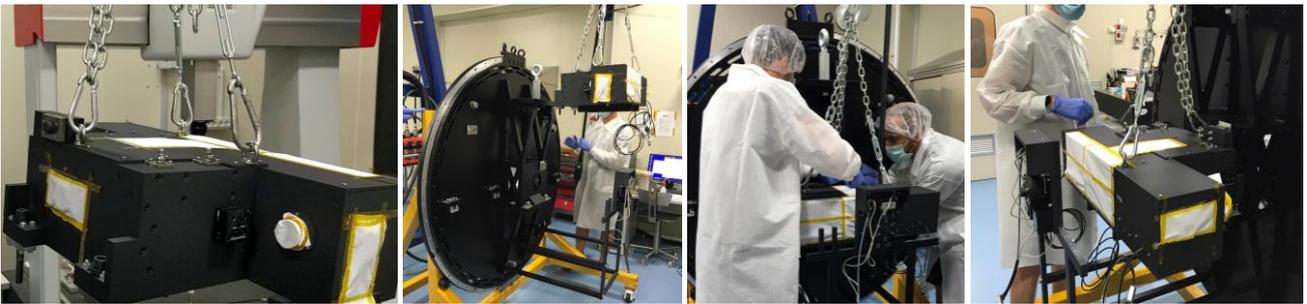

Figure 25. From left to right: CP lifted; CP in front of the IF flange; tightening of the CP KMs; CP installed.

### 5.2 Installation at INAF Padua

The integration of the whole system is performed at INAF Padua. The first steps, shown in Figure 26, consisted in the installation of the IF flange on the NTT simulator and the consequent installation of the CP on the IF flange via the large KMs.

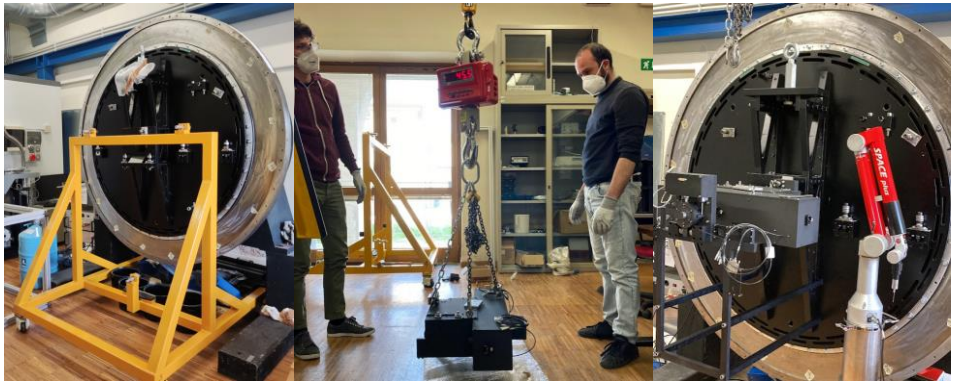

Figure 26. Left: IF flange after the installation on the NTT simulator; the SOXS support is parked in front of the system. Centre and right: installation of the CP.

The next step has been the installation of the Calibration Box and the Acquisition Camera as shown in Figure 27. The first subsystem has been installed with the instrument 'upside-down' while the second one has been installed with the instrument in the 'zero' position.

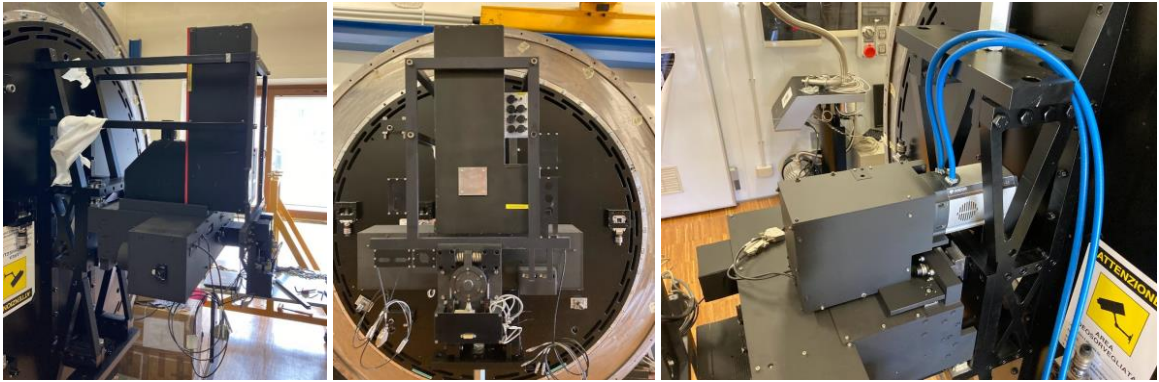

Figure 27. Left and Centre: installation of the CBX. Right: installation of the Acquisition Camera.

In Figure 28 is shown the installation of the corotator. It has been mounted on a dummy support and it has been connected to the cable support. A tracking check has then been successfully performed.

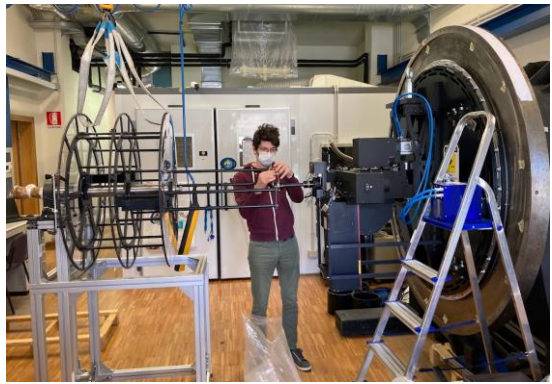

Figure 28. Installation of the corotator and connection to the rest o the system.

Following the installation of the corotator, the cabinets have been installed on the side of the corotator (Figure 29, left) and the flexible rails have been mounted in-between the two (Figure 29, right).

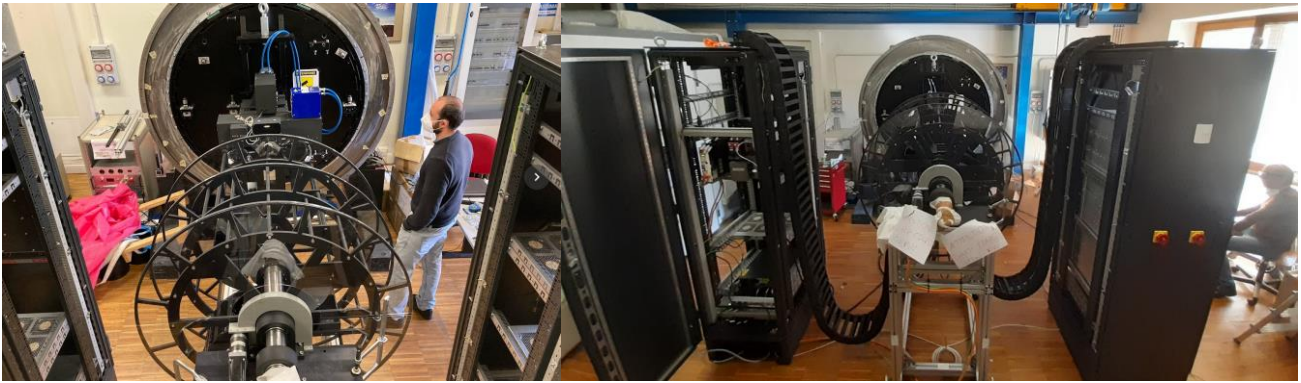

Figure 29. Left: installation of the cabinets. Right: installation of the flexible rails.

The cables connecting CP, Calibration Box, and Acquisition Camera have then been deployed and the routing optimized (see Figure 30: left before, right after).

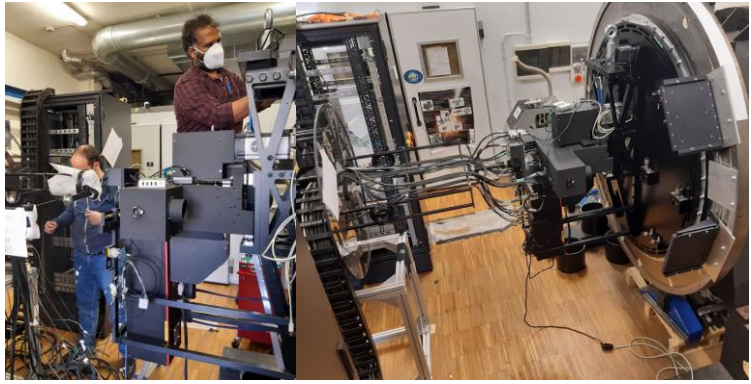

Figure 30. Left: deployment of the cables. Right: routing and tidying up of the cables.

Finally, the UV-VIS spectrograph has been installed and the LN2 lines connected (see Figure 31).

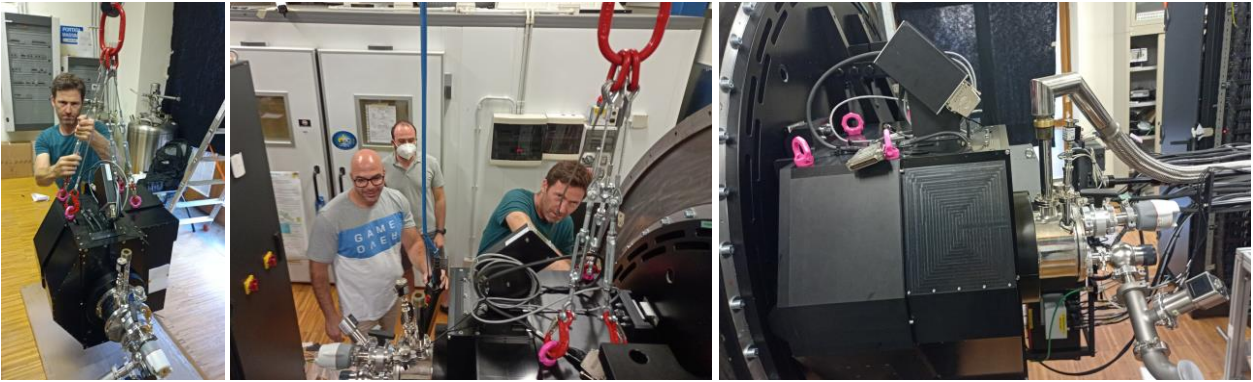

Figure 31. Left: lifting of the UV-VIS spectrograph. Centre: installation on the flange. Right: after the installation with the LN2 line connected.

## 6. CONCLUSIONS

All the mechanical elements of SOXS have been procured except for the platform and the CBX 'v2' and the integration is proceeding. As of today, all the elements except for the NIR spectrograph have been shipped to Padua where the integration of the various subsystems is ongoing.

A summary of the updated work done during the integration of the various subsystem has been presented together with a simplified visual description of their integration in Brera and Padua.

The missing steps to be performed in Brera are the optical alignment and verification of the NIR spectrograph at cold temperatures and its shipment to Padua. All the other tools and subsystems are already in Padua where the final integration is ongoing.